\documentstyle[aps,prl,multicol,amsfonts,amssymb,epsfig]{revtex}
\renewcommand{\v}[1]{\boldsymbol{#1}}

\newcommand{\Ref}[1]{Ref.~\cite{#1}}

\newcommand{\be}{\begin{equation}}
\newcommand{\ee}{\end{equation}}
\newcommand{\ba}{\begin{eqnarray}}
\newcommand{\ea}{\end{eqnarray}}                                   
\newcommand{\nn}{\nonumber} 
\newcommand{\Eq}[1]{Eq.~(\ref{#1})} 

\newcommand{\vev}[1]{\left\langle #1 \right\rangle} 
 
\newcommand{\<}{\langle} 
\renewcommand{\>}{\rangle} 
\newcommand{\Tr}{{\rm Tr}} 
 
\renewcommand{\Im}{{\rm Im}} 
\renewcommand{\Re}{{\rm Re}}

\newcommand{\prt}{\partial} 
\newcommand{\pt}{\partial} 
 
\renewcommand{\t}[1]{{\tilde #1}} 
\newcommand{\up}{\uparrow} 
\newcommand{\down}{\downarrow} 

\newcommand{\ie}{{\it ie~}} 
\newcommand{\eg}{{\it eg~}}

\newcommand{\al}{\alpha} 
\newcommand{\bt}{\beta} 
\newcommand{\del}{\delta} 
 
\newcommand{\eps}{\epsilon} 
 
\newcommand{\ga}{\gamma} 
 
\newcommand{\ka}{\kappa} 
\newcommand{\la}{\lambda}

\renewcommand{\th}{\theta} 
 
\newcommand{\si}{\sigma}

\newcommand{\cA}{ {\cal A} }

\newcommand{\cG}{ {\cal G} }

\newcommand{\cL}{ {\cal L} }

\newcommand{\enu}[1]{ \noindent
\parskip 0pt
 \begin{enumerate} 
 #1 
 \end{enumerate} 
}

\def\journal #1, #2, #3, 1#4#5#6{{\sl #1~}{\bf #2}, #3 (1#4#5#6) }
\def\pr{\journal Phys. Rev., }

\def\prl{\journal Phys. Rev. Lett., }
\def\pl{\journal Phys. Lett., }
\def\np{\journal Nucl. Phys., }

\renewcommand{\v}[1]{{\bbox #1}}

\begin{document}
\draft
\widetext

\title{
Projective Construction
of Non-Abelian Quantum Hall Liquids
}

\author{Xiao-Gang Wen}
\address{
Department of Physics,
Massachusetts Institute of Technology,
Cambridge, MA 02139, USA
}

\maketitle

\widetext
\begin{abstract}
\rightskip 54.8pt

Using projective construction, a generalized parton construction,
we construct many non-Abelian quantum Hall (QH) states,
which include the Pfaffian state at filling fraction $\nu=1/2$.
The projective construction allows us to
calculate the bulk and the edge effective theory for the constructed QH state.
We illustrate how to use the bulk effective theory to calculate the
ground state degeneracy of non-Abelian QH liquids on torus.
We point out that the full description of the effective theory
requires both the effective Lagrangian and the definition of electron
operators. The latter generates all physical states and defines the
gauge structure of the theory.
\end{abstract}


\begin{multicols}{2}

\narrowtext
\section{Introduction}

QH liquids as a fundamentally new state of matter contain a new kind of order
-- the topological order.\cite{Wtop,rev} The different topological orders
in the QH liquids can be divided into two classes.
The topological orders in the first class -- the Abelian topological orders --
are labeled  by $K$-matrix,\cite{Kmat} which were believed to
describe most of the observed QH liquids.
The second class of topological orders -- the non-Abelian topological orders
-- also exists in QH liquids.\cite{MR,Wnab}
The quasi-particle in the non-Abelian QH
states carry non-Abelian statistics, and their edge states cannot be described
by ``edge phonons'' (which is a collection of harmonic oscillators).

There are two ways to construct non-Abelian QH states.
One \cite{MR,BW,WW,NW,RR,CGT}
is through correlation function in conformal field theories (CFT),
and the other \cite{Wnab,BW} is through the parton construction.\cite{J,Wp}
Both constructions allow us to calculate the structures of edge
states.\cite{Wnab,WWH,MiR,RR,CGT}
However, only the parton construction allows us to calculate the
bulk effective theories, which turn out to be Chern-Simons (CS) theories.

In this paper we introduce the projective construction which generalizes the
parton construction.
Using the projective construction,
we can construct
many old and new non-Abelian (and Abelian) QH states, which include both the
$\nu=1$ bosonic Pfaffian state and the $\nu=1/2$ fermionic Pfaffian
state,\cite{MR,Whalf,WW,WWH,MiR,NW,RR,CGT} as well as the $d$-wave paired
state introduced in \Ref{WW,WWH}.
The projective
construction allows us to calculate both the bulk and the edge effective
theories for the constructed QH states.
The bulk effective theories are complete enough to allow us to calculate the
ground state degeneracies on torus.

Using the projective construction, we find the effective theories
for the $\nu=1$
bosonic Pfaffian state and the $\nu=1/2$ fermionic Pfaffian state to be the
$SO(5)_1$ and the $U(1)\times SO(5)_1$ CS theories.
{}From those effective theories, we calculated the
ground state degeneracies for the both states, which are 3 and 6 on torus.
We also calculated the edge effective theories for the two states.
The results agree with the previous results obtained through the wave
functions.\cite{pfDeg,RR,Whalf,WWH,MiR}

The effective theories of the $\nu=1$ bosonic Pfaffian state and the $\nu=1/2$
fermionic Pfaffian state have been obtained before using different approaches.
The $SO(5)_1$ CS effective theory for the $\nu=1$
bosonic Pfaffian state obtained here is formally
different from the $SU(2)_2$ CS effective theory
obtained in \Ref{FNTW}. Despite both theories give three degenerate ground
states
on torus, the meaning of the gauge fields and the coupling to the external
electromagnetic field $A_\mu$ are quite different.
The effective theory for the $\nu=1/2$ fermionic Pfaffian
state obtained in \Ref{FNTW} is very unusual (which cannot be regarded as an
ordinary CS theory). It
is not clear if such an effective theory is equivalent to
our $U(1)\times SO(5)_1$ CS effective theory for the $\nu=1/2$ fermionic
Pfaffian  state.
In particular it is not clear whether
the effective theory in  \Ref{FNTW} reproduces the six degenerate ground
states on torus.
Another form of effective theory -- the non-Abelian Ginzburg-Landau CS theory --
was obtained in \Ref{FNS} to describe the Pfaffian states. Since the
ground state degeneracies were not calculated, the relation between the
effective theories in \Ref{FNS} and the effective theories obtained in this
paper is unclear at the moment.

In section 2 we introduce the projective
construction using the $U(1)_{l}\times
SU(2)_n$ non-Abelian state as an example. In section 3 we use
the projective construction to construct
the $\nu=1$ bosonic Pfaffian state and the $\nu=1/2$ fermionic Pfaffian state.
This allows us to obtain the bulk and the edge effective theories for the two
states. The projective construction also allows us to construct many new
non-Abelian states and calculate their bulk and edge effective theories.
In section 4 we give a general discussion on the projective
construction. In particular we point out
the importance of the electron operators
in defining the effective bulk theory. We illustrate how the discrete gauge
structure in the effective theory
can affect physical quantities, such as the ground state degeneracy.
The projective construction is a very powerful construction which can
be used to construct many different QH states (both Abelian and non-Abelian).
In section 5 we illustrate how to use the projective construction through some
simple examples.

\section{Projective construction and
$U(1)_{l}\times SU(2)_n$ non-Abelian states}
\label{sec:su}

In this section we are going to use the projective construction
to construct the $U(1)_{l}\times SU(2)_n$ non-Abelian states.\cite{Wnab}
We start with the simplest
example and then generalize to more complicated cases.

We start with a non-Abelian state of spin-1 (bosonic) electrons.\cite{BW}
The wave function is given by
\ba
&&\Phi^b(z_1, m_1; z_2, m_2; ...) \nonumber\\
&=& \sum_{\al_1.\bt_1;...} \chi_s(z_1, \al_1;...) \chi_s(z_1,\bt_1;...)
C^{m_1}_{\al_1\bt_1} ... C^{m_N}_{\al_1\bt_N}
\label{Psi}
\ea
where
$m_i=0,\pm1$,
\be
C^0=\pmatrix{ 0 & \frac{1}{\sqrt{2}}\cr \frac{1}{\sqrt{2}} & 0\cr}\
C^{+1}=\pmatrix{ 1 & 0\cr 0 & 0\cr}\
C^{-1}=\pmatrix{ 0 & 0\cr 0 & 1\cr}
\ee
and $\chi_s(z_1, \al_1;...) $ is the wave function of spin-1/2 fermions
with the first Landau level filled by the spin-up and spin-down particles.
One way to see that the above state is a non-Abelian state is to derive its
low energy effective theory.

To construct  the above wave function using the projective construction,
we start with the following {\em free} fermion
wave function for two species (labeled by $a=1,2$) of spin-1/2 partons
$\psi_{a\al}$:
\be
\Phi_{parton}=\chi_s(z^{(1)}_1, \al_1;...) \chi_s(z^{(2)}_1,\bt_1;...)
\label{Psip}
\ee
where $\al_i, \bt_i =\up,\down$ are spin-1/2 indices.
Then we combine the two spin-1/2 partons into a spin-1
electron. In terms of the electron
operator $\Psi_{m}(z)$ and the parton operator $\psi_{a\al}(z)$, the
combination can be expressed as
\be
\Psi_{m}(z)= \psi_{a\al}(z)\psi_{b\bt}(z) \eps_{ab}C^m_{\al\bt}
\label{Psipsi}
\ee
where $a,b=1,2$.
One can easily see that after setting $z^{(1)}_i=z^{(2)}_i=z_i$ and
symmetrizing $\al_i$ and $\bt_i$ using $C^m_{\al_i \bt_i}$,
$\Phi_{parton}$ in Eq. (\ref{Psip})
reduces to $\Phi^b$  Eq. (\ref{Psi}). Or more precisely
\be
\Phi^b(z_1, m_1; z_2, m_2; ...) = \<0|\prod \Psi_{m_i}(z_i) |\Phi_{parton}\>
\label{PhiPhiparton}
\ee
where $|\Phi_{parton}\>$ is the independent parton state described by
$\Phi_{parton}$.

To obtain the effective theory for state $\Phi^b(z_i, m_i)$, we start with
the effective theory for independent partons
\begin{equation}
i \psi^{\dag}_{a\al} \pt_t  \psi_{a\al} + {1
\over 2m}
\psi^{\dag}_{a\al}(\pt_i-i \frac{e}{2} A_i)^2 \psi_{a\al}
\label{p-eff}
\end{equation}
whose ground state is $|\Phi_{parton}\>$.
The effective theory for the state  $\Phi^b$
 is obtained by
combine the two kinds of partons in the above effective theory into
electrons. Notice that the  effective theory for independent partons
contain $SU(2)$ excitations created by $\psi^\dag_{a\al}\tau^l_{ab}\psi_{b\al}$
where $\tau^l$ are Pauli matrices. We will call such a $SU(2)$ the
color $SU(2)$
to distinguish from the $SU(2)$ spin rotation. From Eq. (\ref{Psipsi}) we
see that the electron operator $\Psi_{m}(z)$ is a color $SU(2)$ singlet.
All physical excitations (created by electron operators) are
color singlets. Thus to combine the partons into electrons, we simply need to
remove all the ``colored'' excitation from the parton theory Eq. (\ref{p-eff})
and project into local color singlet sector. The projection can be realized,
at the Lagrangian level, by introducing a $SU(2)$ gauge field which couples to
the current and the density of the color $SU(2)$:
\ba
\cL &=& i \psi^{\dag}_{a\al} (\del_{ab}\pt_t -i (a_0)_{ab}) \psi_{b\al}
\nonumber\\
&& + {1 \over 2m}
\psi^{\dag}_{a\al}(\del_{ab} \pt_i-i \frac{e}{2} A_i  -i (a_i)_{ab})^2
\psi_{b\al}
\label{su-eff}
\ea
Eq. (\ref{su-eff}) is the effective theory for $\Phi^b$.
Only the gauge invariant operators, such as the electron operator $\Psi_m$,
correspond to physical operators. To see that the effective theory Eq.
(\ref{su-eff}) describe a non-Abelian state, we integrate out the parton
fields $\psi_{a\al}$ and obtain an $SU(2)$ Chern-Simons (CS) theory at level
$k=2$:
\begin{equation}
 \frac{k}{4\pi}{\rm Tr}\eps^{\mu\nu\la}
 (a_\mu \prt_{\nu} a_\la + \frac{2i}{3} a_\mu a_\nu a_\la)
\end{equation}
Although the level $k=1$ $SU(2)$ CS theory contain only Abelian statistics,
the level $k>1$ $SU(2)$ CS theory indeed has quasi-particles carrying
non-Abelian statistics.\cite{Witten}

The edge states of the above non-Abelian state can also be
obtained from the
projective construction. For independent partons described by
Eq. (\ref{p-eff}), the edge theory is simply given by free chiral
fermions in $1+1$D:
\be
i \psi^{\dag}_{a\al} (\pt_t-v \pt_x)  \psi_{a\al}
\label{p-edge}
\ee
The above edge theory is also described by the
$U(1)\times SU_{spin}(2)_2\times SU_{color}(2)_2$ Kac-Moody (KM) algebra.
\cite{Af}
The charge associated with the $U(1)$
is just the electric charge of the electrons.
The combination of the partons into electron is again realized by project
into local $SU_{color}(2)$ singlet sector, which can be simply done
by removing the sector generated by the $SU_{color}(2)_2$ KM algebra
from the edge spectrum. Thus the edge states of the non-Abelian state
$\Phi^b$  is described by the $U(1)_1\times SU_{spin}(2)_2$ KM
algebra.

Here we have assigned a level 1 to the $U(1)$ KM algebra.
The level characterizes how the $U(1)$ charge is quantized.
The definition of the level is the following.
We know that the edge theory contain electron operators
$\Psi_m$. The operators that create quasi-particles are the operators which are
local respect to the electron operators (\ie their correlation with the electron
operators are single valued). Let $\psi_{U(1)}$ be the $SU_{spin}(2)$ singlet
quasi-particle operator which carries the minimum (but non-zero) $U(1)$
charge. The correlation of $\psi_{U(1)}$ has a form $\vev{ \psi_{U(1)}^\dag(x)
\psi_{U(1)}(0)} \sim 1/x^{h}$. Then the level of the $U(1)$ KM algebra
is defined as $l=1/h$. (According to this definition, the $U(1)$ KM algebra
that describes the edge excitations of the $\nu=1/m$ Laughlin state
can be more specifically denoted as $U(1)_m$ KM algebra.)
Since its edge state is described by the $U(1)_1\times SU_{spin}(2)_2$ KM
algebra, we will call the non-Abelian state $\Phi^b$ the
$U(1)_1\times SU_{spin}(2)_2$ state. Note that the $U(1)_1\times
SU_{spin}(2)_2$ non-Abelian state can have a bulk effective theory which is
a purely $SU_{color}(2)_2$ CS theory.

A slightly more complicated non-Abelian state of spin-1 fermionic electrons
is given by
\ba
&&\Phi^f(z_1, m_1; z_2, m_2; ...)
= \prod (z_i-z_j) \times \nonumber\\
&&\sum_{\al_1.\bt_1;...} \chi_s(z_1, \al_1;...) \chi_s(z_1,\bt_1;...)
C^{m_1}_{\al_1\bt_1} ... C^{m_N}_{\al_1\bt_N}
\label{Psi-f}
\ea
To obtain the projective construction, we need to split the electron
into one charge $e/2$ parton $\psi_0$ and two charge $e/4$ partons
$\psi_a|_a=1,2$:
\be
\Psi_{m}(z)= \psi_0(z)\psi_{a\al}(z)\psi_{b\bt}(z) \eps_{ab}C^m_{\al\bt}
\label{Psipsi-f}
\ee
Following the similar arguments used above, we obtain the effective
theory for $\Phi^f$:
\begin{eqnarray}
&& i \psi^{\dag}_{0} (\pt_t + 2 i b_0)
\psi_{0}
+ {1 \over 2m}
\psi^{\dag}_{0}( \pt_i-i \frac{e}{2} A_i  + 2 i b_0)^2 \psi_{0}
\nonumber \\
&& +i \psi^{\dag}_{a\al} (\del_{ab}\pt_t -i (a_0)_{ab}-i b_0\del_{ab})
\psi_{b\al}  \nonumber\\
&&
+ {1 \over 2m} \psi^{\dag}_{a\al}
( \pt_i-i \frac{e}{4} A_i -ia_i - i b_0)^2_{ab}
\psi_{b\al}
\label{su-eff-f}
\end{eqnarray}
Here an extra $U(1)$ gauge field $b_\mu$ is introduced to combine $\psi_0$ and
$\psi_{1,2}$ together. Note that the electron operator $\Psi_m$ carries no
$b_\mu$ charge. After integrating out the parton fields, the effective theory
becomes the $U(1)\times SU(2)_2$ CS theory.
The edge states for independent partons are described by the
$U(1)\times U(1)\times SU_{spin}(2)_2\times SU_{color}(2)_2$ KM algebra.
After the projection,
the edge states for the fermion non-Abelian state are described by the
\begin{eqnarray}
&&\frac{U(1)\times U(1)\times SU_{spin}(2)_2\times SU_{color}(2)_2}{U(1)\times
SU_{color}(2)_2} \nonumber\\
&=&U(1)_2\times SU_{spin}(2)_2
\end{eqnarray}
KM algebra.

More general non-Abelian state of spin-$\frac{n}{2}$ electrons
is given by
\ba
&&\Phi^{(n,k)}(z_1, m_1; z_2, m_2; ...)
= \prod (z_i-z_j)^k \times \nonumber\\
&&\sum_{\al_{ai}}
C^{m_1}_{\al_{11}..\al_{n1}} ... C^{m_N}_{\al_{1N}..\al_{nN}} \times \nonumber\\
&&\chi_s(z_1, \al_{11};...;z_N, \al_{1N})...
\chi_s(z_1,\al_{n1};...;z_N,\al_{nN})
\label{Psi-fn}
\ea
where $C^m_{\al_1,...,\al_n}$ form a basis of rank-$n$ symmetric tensors, and
$m$ is the quantum number of the total spin $S_z$ for each electron.
In the bulk, the above state is described by the
$(U(1))^k\times SU_{color}(n)_2$
effective CS theory. The edge excitations are described by the
$U(1)_{k+\frac{n}{2}} \times SU_{spin}(2)_n$ KM algebra.
Such a state will be called $U(1)_{k+\frac{n}{2}} \times SU_{spin}(2)_n$
non-Abelian state.

The above result provides an example that the group for the edge
KM algebra,
$U(1)_{k+\frac{n}{2}} \times SU_{spin}(2)_n$, and the group for the bulk
CS effective theory,  $(U(1))^k\times SU_{color}(n)_2$,  can be quite different.

\section{Projective construction and the effective theory of Pfaffian state}
\label{sec:pf}

Using the  $U(1)_{k+\frac{n}{2}} \times SU_{spin}(2)_n$
non-Abelian states, we can construct
new types of non-Abelian states. Let us start with the simplest
$U(1)_1\times SU_{spin}(2)_2$ state.

To construct a new non-Abelian state from the $U(1)_1\times SU_{spin}(2)_2$
state,  we simply make a further local projection $S_z=0$,
in addition to the local color singlet projection.
This $S_z=0$ projection can be realized by identifying
\be
\Psi_e\equiv \frac{1}{\sqrt{2}} \Psi_{m=0}
\label{Psiepsi}
\ee
as the only physical electron operator. $\Psi_{m=\pm 1}$ are
regarded as unphysical since $S_z\neq 0$. The physical Hilbert space
is generated by $\Psi_{m=0}$ only. Therefore after the $S_z=0$ projection,
the wave function of the new non-Abelian state is given by (see
\Eq{PhiPhiparton})
\begin{equation}
\Phi^{pf}(z_1,..., z_N) =\<0| \prod_i \Psi_e(z_i) |\Phi_{parton}\>
\label{pf-psichi}
\end{equation}
which can be regarded as a wave function of spinless electrons.

The projective construction allows us to obtain the low energy effective
theory for the above new non-Abelian state.
At Lagrangian level, the $S_z=0$
projection can be realized by introducing an extra
$U_{S_z}(1)$ gauge field $c_\mu$
that couples to $S_z$. This yield the effective theory
for the new non-Abelian state:
\ba
&&
i \psi^{\dag}_{a\al} (\del_{ab}\del_{\al \bt}\pt_t -i \del_{\al \bt}(a_0)_{ab}
-i c_0 \si^3_{\al\bt} \del_{ab} )\psi_{b\bt}  \nonumber\\
&& + {1 \over 2m}
\psi^{\dag}_{a\al}(\pt_i-i \frac{e}{2} A_i  -i a_i -i c_0
\si^3)^2_{a\al,b\bt} \psi_{b\bt}
\label{pf-eff}
\ea
After integrating out the fermions, we get a
$U_{S_Z}(1)\times SU_{color}(2)_2$
CS theory.

The edge excitations of the $\Phi^{pf}$ state can also be obtained through the
projective construction.
Since the  $\Phi^{pf}$ state is obtained from the $U(1)_1\times SU_{spin}(2)_2$
state by making an additional
local $S_z=0$ projection, thus the edge states of the
$\Phi^{pf}$ state can also be obtain from that of the  $U(1)_1\times
SU_{spin}(2)_2$ state by making a local $S_z=0$ projection.
Note that the edge excitations that correspond to the $S_z$ fluctuations
is described by the $U_{S_z}(1)$ KM algebra generated by the $S_z$ current.
Thus the edge excitations of the $\Phi^{pf}$ state is described by the
$U(1)_1\times (SU_{spin}(2)_2/U_{S_z}(1))$ coset theory.\cite{coset,Wp}
Since the
$SU_{spin}(2)_2/U_{S_z}(1)$ coset theory is nothing but an $Ising$ model
(or a Majorana fermion theory) with the central charge $c=1/2$,\cite{GQ}
the edge
theory for new non-Abelian state
can also be denoted as $U(1)_1\times Ising$ theory.

The second
way to obtain the edge theory is to note that all edge excitations
are generated by $\Psi_e$ in Eq. (\ref{Psiepsi}) and $\Psi_e^\dag$.
Thus we can use the algebra of $(\Psi_e, \Psi_e^\dag)$ to describe
the edge excitations. The algebra of  $(\Psi_e,\Psi_e^\dag)$ can be obtained
from their operator product expansion (OPE), which can be calculated easily
since $\Psi_e$ can be expressed as a product of free chiral fermion operators
Eq. (\ref{Psiepsi}). Using the OPE for free chiral fermions
$\psi(z)\psi^\dag(0)=1/z$,
we find the following closed OPE generated by $(\Psi_e,\Psi_e^\dag)$:
\ba
\Psi_e^\dag(z)\Psi_e(0) &=& \frac{1}{z^2} + \frac{J(0)}{z} + O(z^0) \nn\\
\Psi_e(z)\Psi_e^\dag(0) &=& \frac{1}{z^2} - \frac{J(0)}{z} + O(z^0) \nn\\
J(z) \Psi_e(0) &=& -    \frac{\Psi_e(0)}{z} + O(z^0) \nn\\
J(z) \Psi_e^\dag(0) &=& \frac{\Psi_e(0)}{z} + O(z^0) \nonumber\\
J(z)J(0) &=& \frac{1}{z^2} + O(z^2)
\label{pwOPE}
\ea
It is not hard to see that
\be
J=\frac{1}{2} \psi_{a\al}^\dag \psi_{a\al}
\label{Jphi}
\ee
To obtain the Hilbert space generated by $(\Psi_e,\Psi_e^\dag)$ (or
equivalently, to find the representation of the above OPE \Eq{pwOPE}),
we note that
in addition to the representation Eq. (\ref{Psiepsi}), the following
reprensentation of $\Psi_e$
\be
\Psi_e= \psi\eta,\ \ \  \Psi_e^\dag = \eta\psi^\dag
\label{Psieeta}
\ee
also reproduces the exactly the same OPE Eq. (\ref{pwOPE}). (Here
$\psi$ is a free chiral fermion,
$\psi(z)\psi^\dag(0)=1/z$, and $\eta$ a Majorana fermion,
$\eta(z)\eta(0)=1/z$.) Thus $\Psi_{m=0}$ and $\psi\eta$ have exactly
the same correlations, and we can identify $\Psi_{m=0}= \psi\eta$.
Since the electron operator $\Psi_e$ can be expressed as a product
of a free chiral fermion $\psi$ in the $U(1)$ theory and a Majorana fermion $\eta$
in the $Ising$ theory, and since all physical edge excitations are generated by
electron operators, the edge theory of the new non-Abelian state is described
by the $U(1)\times Ising$ CFT theory.

To obtain an explicit expression of the wave function for our new non-Abelian
state, let us first review
a relation between the edge theory and the bulk wave function.\cite{WWH}
As we mentioned above that for independent partons, the edge state is described
by free chiral fermions $\psi_{a\al}$ (see Eq. (\ref{p-edge})). This edge
theory and the independent-parton wave function Eq. (\ref{Psip}) are closely
related. As pointed out in \Ref{MR}, the following correlation in the 1+1D
free chiral fermion theory
\be
\vev {e^{-iN\phi(z_\infty)} \prod_{i=1..N} \psi(z_i) } \sim \prod_{ij} (z_i-z_j)
\ee
is proportional to the analytic part of the spinless
electron wave function of filled
first Landau level, $\prod_{ij} (z_i-z_j) e^{-\sum|z_i|^2/4}$. Here $\psi$ is
a free chiral fermion field, $z$ is given by
$z=x-v t=x+iv \tau$ for complex time
$\tau=it$, and $e^{i\phi(z)} =\psi(z)$ is the bosonized form of the free
chiral fermion operator. One can show, through bosonization,
that $\frac{1}{2\pi} \pt_x \phi =\psi^\dag(x)\psi(x)$.

Generalizing the above relation, we find that
\ba
&&\vev{e^{-\frac{1}{2}iN\phi(z_\infty)} \prod_{i=1..N}\psi_{1,\al_i}(z^{(1)}_i)
\psi_{2,\bt_i}(z^{(2)}_i) } \nonumber\\
&\sim&
\chi_{an}(z^{(1)}_1, \al_1;...) \chi_{an}(z^{(2)}_1,\bt_1;...)
\label{p-psichi}
\ea
where $\chi_{an}(z_1, \al_1;...)$ is the analytic part of
$\chi_s(z_1, \al_1;...)$, and
$\frac{1}{2\pi} \pt_x \phi = \psi^\dag_{a\al}(x)\psi_{a\al}(x)$,
which is the total density
operator of the fermions $\psi_{a\al}$.
We see that (the analytic part of) the
independent-parton wave function can be expressed as a
correlation of the independent parton operators.

Similarly, after combining the partons into electrons, the  wave function
$\Phi^b$ for
the $U(1)_1\times SU_{spin}(2)_2$ non-Abelian
state can be expressed as a
correlation of the electron operator.
Actually from Eq. (\ref{Psipsi}) and Eq. (\ref{p-psichi}), we see that
\be
\vev {e^{-\frac{1}{2}iN\phi(z_\infty)} \prod_{i=1..N}\Psi_{m_i}(z_i)
} \sim
\Phi^b_{an}(z_1, m_1;...)
\label{su-psichi}
\ee
where $\Phi^b_{an}(z_1, m_1;...)$ is the analytic part of the $\Phi^b$ in Eq.
(\ref{Psi}). Note that both $e^{-2iN\phi}$ and $\Psi_{m_i}$ are
$SU_{color}(2)$ singlets. Thus they are operators in the projected
$U(1)_1\times SU_{spin}(2)_2$ theory.

Now it is clear that the wave function of the new non-Abelian state
$\Phi^{pf}$ can also be expressed as a correlation.
The analytic part of $\Phi^{pf}$ is
\be
\vev {e^{-\frac{1}{2}iN\phi(z_\infty)} \prod_{i=1..N}\Psi_e(z_i)
} \sim
\Phi^{pf}_{an}(z_1,..., z_N)
\label{pf-psichi1}
\ee
Since $\Psi_e(z)$ is simply a product of free fermion operators,
the expression \Eq{pf-psichi1} can help us the calculate the wave function.
Note that $\Psi_e(z)$ can also be expressed as
$\Psi_e= \psi\eta$.
This allows us to calculated the wave function
of our non-Abelian state more easily:
\ba
&&\Phi^{pf}_{an}(z_1,..., z_N)  \propto
\vev {e^{-iN\phi_e(z_\infty)} \prod_{i=1..N}\Psi_e(z_i)
} \nn\\
&=& \cA \left(\frac{1}{z_1-z_2} \frac{1}{z_3-z_4}...\right)
\prod_{\vev{ij}} (z_i-z_j)
\label{pf-Psie}
\ea
where $\cA$ is the total anti-symmetrization operator. The first factor
$\cA \left(\frac{1}{z_1-z_2} \frac{1}{z_3-z_4}...\right) $ comes from the
$\eta$ correlation and the second one $ \prod_{\vev{ij}} (z_i-z_j)$ from
the $\psi$ correlation.
From the explicit form of the wave function, we find that our new non-Abelian
state is actually
not new. It is nothing but the Pfaffian state (for bosonic electrons)
first introduced by Moore and Read \cite{MR}.

However, we did get some new results. Let us summarize the new results here:
\enu{
\item
Using an algebraic method, we find that the Pfaffian state can be obtained from
the projective construction.
\item
The projective construction allows us to derive the bulk low energy
effective theory Eq. (\ref{pf-eff}), which is
a $U_{S_z}(1)\times SU_{color}(2)_2$ CS theory.
\item
The projective construction also allows us to obtain the edge effective theory,
the $U(1)_1\times Ising$ CFT theory. This agrees with the old results
obtained using different methods.\cite{Whalf,WWH,MiR,RR,CGT}
}

The projective construction
can also be applied to other (non-)Abelian states, which
allows us to generate many new non-Abelian states. In the following,
as examples, we will only
give the final results of several other non-Abelian states obtained using the
projective construction.

Making the $S_z=0$ projection for the
$U(1)_2\times SU_{spin}(2)_2$ non-Abelian state, we obtain the following
results:
\enu{
\item
The Pfaffian state for (fermionic) eletrons can be obtained from the
projective construction.
Let $\Psi_e$ to be $\Psi_{m=0}$ in Eq. (\ref{Psipsi-f}),
or
\be
\Psi_e(z)= \psi_0(z)(\psi_{a1}(z)\psi_{b2}(z) +
\psi_{a2}(z)\psi_{b1}(z) )\eps_{ab},
\label{Psipsi-fpf}
\ee
then the wave function of the fermionic Pfaffian state can be expressed as
\ba
&&\vev {e^{-iN\phi_e(z_\infty)} \prod_{i=1..N}\Psi_e(z_i) } \nonumber\\
&\propto& \cA \left(\frac{1}{z_1-z_2} \frac{1}{z_3-z_4}...\right)
\prod_{\vev{ij}} (z_i-z_j)^2
\label{fpf-Psie}
\ea
The above result implies that the fermionic Pfaffian wave function can be
obtained
from the wave function $\prod (z_i-z_j) \chi_s^2$
of the $U(1)_2\times SU_{spin}(2)_2$
state through the $S_z=0$ projection.
\item
The bulk low energy
effective theory for the fermionic Pfaffian state is given by
\begin{eqnarray}
&& i \psi^{\dag}_{0} (\pt_t + 2 i b_0) \psi_{0}
+ {1 \over 2m}
\psi^{\dag}_{0}( \pt_i-i \frac{e}{2} A_i  + 2 i b_0)^2 \psi_{0} \nonumber\\
&&
+i \psi^{\dag}_{a\al} (\del_{\al\bt}\del_{ab}(\pt_t -i b_0)
-i \del_{\al\bt}(a_0)_{ab}-i
c_0\del_{ab}\si_{\al\bt}^3)
\psi_{b\bt}
\nonumber \\ &&
+ {1 \over 2m} \psi^{\dag}_{a\al}
( \pt_i-i \frac{e}{4} A_i -ia_i - i b_i-ic_i\si^3)^2_{a\al,b\bt}
\psi_{b\al}
\label{su-eff-fpf}
\end{eqnarray}
Comparing to Eq. (\ref{su-eff-f}), we introduced an additional $U_{S_z}(1)$
gauge field $c_\mu$ to
perform the $S_z=0$ projection. After integrating out the fermions, we obtain
the $U(1)\times U_{S_z}(1)\times SU_{color}(2)_2$ CS theory.
This effective theory, at least formally,
is quite different from another effective theory
obtained in \Ref{FNTW} for the same fermionic Pfaffian state.
\item
The edge effective theory obtained from the projective construction
is the
\begin{eqnarray}
&&\frac{ U(1)\times U(1)\times SU_{spin}(2)_2\times SU_{color}(2)_2}{U(1)
\times U_{S_z}(1)\times SU_{color}(2)_2} \nonumber\\
&=& U(1)_2\times  Ising
\end{eqnarray}
KM algebra.
}

Making the $S_z=0$ projection for the
$U(1)_{k+\frac{n}{2}} \times SU_{spin}(2)_{n={\rm even}}$ state
(with wave function $\Phi^{(n,k)}$ in
Eq. (\ref{Psi-fn})), we obtain the following
results:
\enu{
\item
The bulk low energy
effective theory of the constructed state is given by
$(U(1))^{k}\times U_{S_z}(1) \times SU_{color}(n)_2$ CS theory
\item
The edge effective theory obtained from the projective construction
is the
$U(1)_{k+\frac{n}{2}} \times \left(SU_{spin}(2)_n/U_{S_z}(1) \right)
=U(1)_{k+\frac{n}{2}}\times PF_n$ CFT theory. Here $PF_n$ is the $Z_n$
parafermion theory.\cite{pfermion} Note that $Ising=PF_2$.
\item
The electron operator is given by
$\Psi_e =\la e^{i\ga \phi}$ in the $U(1)_{k+\frac{n}{2}}\times PF_n$ theory,
where $\la$ is the $Z_n$
parafermion current operator $\psi_{n/2}$(see \Ref{pfermion}, \Ref{GQ}
and \Ref{BW}).
Therefore the wave function for the
constructed non-Abelian state is a correlation function of the parafermions
times $\prod_{ij}(z_i-z_j)^{\ga^2}$.
We would like to point out that this state is not the parafermion non-Abelian
state studied in \Ref{RRpF} when $n\neq 2$. The latter is constructed using
parafermion current operator $\psi_{1}$.
}

We note that when $n=4$, the $\psi_2$ has the following OPE
\begin{equation}
\psi_2(z)\psi_2(0)\sim \frac{1}{z^2}
\end{equation}
Thus the above $U(1)_{k+2}\times PF_4$ state is just the $d$-wave paired
state introduced in \Ref{WW}.

\section{Projective construction -- a general discussion}

The low energy effective theory of QH liquid,
for example the one for the bosonic Pfaffian state
Eq. (\ref{pf-eff}), has a finite energy gap for all its excitations (on a
space with no boundary). Thus naively one might expect the low energy
effective theories for QH liquids are trivial since there are simply no low
energy excitations. Certainly this point of view is incorrect. The effective
theories for QH liquids have non-trivial ground state degeneracies which depend
on the topology of the space.\cite{Wtop}
Such theories are called topological theories.\cite{Witten}
Different QH liquids (or topological orders) are described by different
topological theories.
Thus we can say that the topological orders in QH liquids are characterized
by topological theories, just like symmetry broken phases are characterized by
Ginzburg-Landau theories.
In many cases, a topological theories can take many different forms. Thus to
know whether
two topological theories are equivalent or not, it is important to
compare their physical properties, such as the ground state degeneracies.
It is those physical properties that define a topological theory.

One important
issue is that given a QH liquid, how to derive its effective topological
theory which describe the topological order in the QH liquid. From the
discussions in the above sections,
we see that if a QH liquid can be obtained through the projective construction,
then there is a way to calculate its
effective topological theory. In the following, we will give a general
discussion of the projective construction.

One starts with a few parton fields $\psi_a$ (where $a=1,...,n$),
each with electric charge $Q_a$.
Thus for independent parton model, the effective theory is
\begin{equation}
\cL_{eff}=i \psi^{\dag}_{a} \pt_t  \psi_{a}
+ {1 \over 2m} \psi^{\dag}_{a} (\pt_i -i Q_a A_i)^2 \psi_{a}
\label{eff0-g}
\end{equation}
However, the Hilbert space generated by the parton fields
$(\psi_a,\psi_a^\dag)$ is simply too big. The physical Hilbert space,
generated by electron operators $(\Psi_e,\Psi_e^\dag)$ and the electrical
current operators $(J_0, J_i)=(\sum_a \psi_a^\dag Q_a \psi_a,
\Im \sum_a \psi_a^\dag Q_a \prt_i \psi_a)$,
is a subspace of the parton Hilbert space.  Thus it is extremely
important the give the definition of electron operators, in order to even
define the theory.
In general, there can be several electron operators. Here for simplicity we
will only consider the case with one electron operator, which takes the form
\be
\Psi_e= \sum_m C_m \prod_a \psi_{a}^{n_a^{(m)}}(z)
\label{PsiEpsi-g}
\ee
where $n_a^{(m)} =0,1$.
The total charge of the electron operator is $e$, hence
\be
Q_a n_a^{(m)}=e
\ee
for any $m$.
Since the physical Hilbert space is generated only by the electron operators
and the electrical current operators,
our model is actually a gauge theory. Let $\cG$ be the group of
transformations on the parton fields $\psi_a \to W_{ab}\psi_b$ that leave
the electron operator and the  electrical current operators
unchanged. By definition
\be
W^\dag Q W =Q,\ \ \ W\in \cG
\label{WQ}
\ee
where $Q$ is a diagonal matrix with diagonal elements $(Q_1,Q_2,...)$.
Such a matrix is denoted as ${\rm diag}(Q_1,Q_2,...)$.
Note that the electron operator is invariant
even under a local transformation
\be
\psi_a(\v x) \to W_{ab}(\v x) \psi_b(\v x) , \ \ \  W_{ab}(\v x) \in \cG
\label{gaugeT}
\ee
Because all the physical states are generated by the electron operators
and the electrical current operators, the
transformation $W_{ab}(\v x)$ is actually a gauge transformation. To realize the
gauge structure (\ie to project onto the physical Hilbert space), we need to
include gauge fields in Eq. (\ref{eff0-g}) so that it has a proper gauge
invariance.

In general the gauge group $\cG$ can contain several disconnected pieces.
Let $\cG_c$ be the connected piece of $\cG$ which contain the identity
(and $\cG_c$ itself is a subgroup of $\cG$).
Then the gauge structure associated with $\cG_c$ can be realized through
gauge fields $a_\mu(\v x)$ which take value in the Lie algebra of $\cG_c$,
$L_{\cG_c}$:
$a_\mu(\v x) \in L_{\cG_c}$. (\eg if $\cG_c=SU(n)$, then $a_\mu(\v x)$ are
traceless Hermitian matrices.) After including the gauge fields,
the parton theory becomes
\begin{equation}
i \psi^{\dag}_{a} (\del_{ab} \pt_t  -i (a_0)_{ab})\psi_{b}
+ {1 \over 2m} \psi^{\dag}_{a} (\pt_i -i Q A_i -i a_i)^2_{ab} \psi_{b}
\label{eff-nab-g}
\end{equation}
The above Lagrangian
is the low energy effective theory of the QH liquid.
We would like to stress that the Eq. (\ref{eff-nab-g}) alone does not provide
a complete description of the QH liquid. In particular, Eq. (\ref{eff-nab-g})
only includes the gauge structure associated with $\cG_c$. {\em Only  Eq.
(\ref{eff-nab-g}) together with the definition of electron operators
Eq. (\ref{PsiEpsi-g}) provide a complete description of the QH liquid}.
The invariance of the electron
operators (and the electrical current operators)
gives rise to the full gauge group $\cG$ which may contain
discrete gauge transformations, in addition to the
continuous transformation $\cG_c$ described by the gauge fields $a_\mu$.
We will see later that
the discrete gauge transformations are important and can affect physical
properties of the theory, such as the ground state degeneracies.

There is an important issue that we have overlooked in the above discussion.
The Langrangian in general may not describe a state with finite energy gap.
One way to get a state with finite gap is to assume each kind of partons
form an integral QH state with filling fraction $\nu_a=m_a$. In the following
we will examine when this assumption can be self consistent.
Under the assumption $\nu_a=m_a$, we can integrating out the parton fields and
get a CS theory
\ba
\cL&=& \frac{1}{4\pi}\eps^{\mu\nu\la}\Tr (M a_\mu \prt_\nu a_\la)
+ \frac{1}{2\pi}\eps^{\mu\nu\la}A_\mu \Tr (M Q \prt_\nu a_\la) \nonumber\\
&& + \frac{1}{4\pi} \eps^{\mu\nu\la}A_\mu \prt_\nu A_\la \Tr (M Q^2 )
\label{CS-g}
\ea
where $M={\rm diag}(m_1,m_2,...)$.
Here we have assumed that there is no gauge symmetry breaking (or no Higgs
mechanism). Therefore $M$ and $M Q$ must be invariant under the gauge
transformation $\cG$, which requires
\be
W^\dag M W =M,\ \ \
W \in \cG
\label{WM}
\ee

The Eq. (\ref{CS-g}) is obtained in the following way. First we assume $a_\mu$
to be diagonal, which can be regarded as gauge fields for the maximum Abelian
subgroup of $\cG_c$. In this case one can obtain Eq. (\ref{CS-g}) easily.
Since we assume there is no gauge symmetry breaking, the effective theory has
full $\cG_c$ gauge invariance. This allows us to show that Eq. (\ref{CS-g})
to be valid for generic $a_\mu$ in $\cG_c$.

The equation of motion $\prt \cL/\prt a_0$ leads to a solution $\bar a_i$
which can be chosen to be diagonal: $\bar a_\i={\rm diag}(\bar a^{(1)}_i,
\bar a^{(2)}_i,...)$. We note that $\bar a^{(a)}_i$ are proportional to $A_i$:
$\bar a^{(a)}_i = f_a A_i$. If we shift $a_\mu$ to $\t a_\mu=a_\mu+ FA_i$
where $F={\rm diag}(f_1, f_2, ...)$, then the equation of motion will give us
$\t a_i=0$. The shift changes $Q_a$. We see that one can redefine
$Q_a$ through a shift of $a_\mu$ to make $\bar a_i=0$.
In the following we will assume that $Q_a$ are chosen such that $\bar a_i=0$.
This requires
\begin{equation}
 \Tr(t M Q)=0
\label{tMQ}
\end{equation}
for any matrix $t$ in the Lie algebra of $\cG_c$.
From $\prt \cL/\prt (a_0)_{aa}=\rho_a$, we get the
density of the $a^{th}$ parton:
\be
\rho_a =
\frac{1}{2\pi} m_a Q_a  \prt_i A_j\eps^{ij}
\ee
Since $Q_a A_i$
happen to be the total gauge field seen by the $a^{th}$ parton, thus the
$a^{th}$ parton always has a filling fraction $m_a$ regardless how we choose
$m_a$. Here we only require that
$m_a$ are chosen such that $M$ satisfies Eq. (\ref{WM}), it leads to
$a_i=0$ as a solution to the  equation of motion, and
$\rho_a$ are all positive.

The Eq. (\ref{eff-nab-g}), Eq. (\ref{PsiEpsi-g}), Eq. (\ref{WQ}),
\Eq{WM}, and \Eq{tMQ}
form a complete description of the QH liquids. One can calculate
all physical properties, such as the ground state degeneracies,
of the QH liquids from those equations.

In the above discussion we only have one electron operators. In general, there
can be several electron operators,
and the above discussion can be generalize in a
straight forward way to cover those more general cases.
For example the gauge group $\cG$ is formed by transformations that leave
all the electron operators invariant.

After setting up the bulk effective theory, it is easy to obtain the edge
effective theory. For independent partons the edge states contain
$n_{edge}=\sum_a |m_a|$ branches. Each branch is described by a
free chiral fermion theory or a $U(1)$ KM algebra.
Thus the edge effective theory for the independent partons is given by
\be
\cL_{edge}= i\psi^\dag_{al} (\prt_t-v_a \prt_x) \psi_{al}
\label{Ledge}
\ee
where $l=1,...,|m_a|$ and $v_a$ has the same sign as $m_a$. The above theory
is denoted as the $U^{n_{edge}}(1)$ theory. The true edge effective theory
for the physical states is obtained through the coset construction
\cite{coset} as
the $U^{n_{edge}}(1)/\cG$ coset theory. Note that we not only need to
remove excitations associated with the $\cG$ KM algebra (which give us the
$U^{n_{edge}}(1)/\cG_c$ coset theory), we also need to require the
physical states to be invariant under all the discrete gauge transformations
in $\cG$. Another way (which is conceptually better) to get the edge theory
is to setup the OPE algebra of the electron and the current operators, and
generate the edge states through the algebra.

In the following we will outline how to calculate ground state degeneracy on
torus.
From \Ref{WZ}, we see that to get the  ground state degeneracy on torus
we may reduce the non-Abelian gauge fields to Abelian ones, \ie to reduce
the gauge group $\cG_c$ to the maximum Abelian subgroup $\cG_{abl}$ which is
formed by diagonal matrices.
The Abelian version of the effective Lagrangian has a form
\begin{equation}
i \psi^{\dag}_{a} (\pt_t  -i a^I p^I_a )\psi_{a}
+ {1 \over 2m} \psi^{\dag}_{a} (\pt_i -i Q_a A_i
-i a^I_i q^I_a)^2 \psi_{a}
\label{eff-g}
\end{equation}
where $I=1,..., \ka$. The electron operator is given by Eq. (\ref{PsiEpsi-g}).
The gauge invariance of the electron operator requires
\be
n_a^{(m)} p^I_a =0
\ee
for any $m$ and $I$ (see \Eq{PsiEpsi-g}).
In addition to the Abelian gauge structure described by $a^I_\mu$, there are
also discrete gauge transformation generated by $W_i\in \cG$ which
leave the Abelian subgroup unchanged:
\be
W_i^\dag \cG_{abl} W_i = \cG_{abl}
\label{WGW}
\ee
The above $W_i$'s form a discrete group. The
Eq. (\ref{WGW}) can be reduced to
the following matrix equation:
$W_i$ is a discrete gauge transformation if and only if
$W_i \in \cG$ and
there exists a $\ka\times \ka$ matrix $T_i$ such that
\be
\sum_a (W_i^\dag)_{ba}p^I_a (W_i)_{ac}(T_i)_{IJ} = p^J_b \del_{bc}
\label{MpW}
\ee
The Lagrangian in Eq. (\ref{eff-g}) and the electron operator in
Eq. (\ref{PsiEpsi-g}) are invariant under
the discrete gauge transformation
\be
\psi_a \to (W_i)_{ab} \psi_{b}
\label{Wpsi}
\ee
\be
a^I_\mu \to (T_i)_{IJ} a^J_\mu
\label{Ma}
\ee

After integrating out the parton fields from Eq. (\ref{eff-g}), we
obtain a $U(1)$ CS effective theory
\be
\frac{\t K_{IJ}}{4\pi} a_{I\mu}\pt_\nu a_{J\la} \eps_{\mu\nu\la}
+\frac{q_I}{2\pi} A_{\mu}\pt_\nu a_{I\la} \eps_{\mu\nu\la}
+\frac{\nu e^2}{4\pi} A_{\mu}\pt_\nu A_{\la} \eps_{\mu\nu\la}
\label{LeffK}
\ee
where
\ba
\t K_{IJ} &=& \sum_a m_a p^I_a p^J_a \nonumber\\
q_I &=& \sum_a m_a Q_a p^I_a \nonumber\\
\nu &=& \frac{\sum_a m_a Q_a Q_a}{e^2}
\label{Kqnu}
\ea
Since we require $a_i=0$ to be a solution to the equation of motion,
$m_a$ must be chosen to satisfy
\be
\sum_a m_a Q_a p^I_a =q_I=0
\label{mQp}
\ee
for all $I$.  (Note that \Eq{mQp} is just a special case of \Eq{tMQ}.)
$m_a$ should also satisfy \Eq{WM}.  For such $m_a$,
$\nu$ becomes the total filling fraction of the QH liquid.

Now let us use the Abelian version of the effective theory to calculate
the ground state degeneracy on torus.
Following \Ref{latt,WZ}, the low energy degrees of freedom are described
by $\v u$ and $\v v$:
\begin{equation}
a^I_1(x_1,x_2,t)=2\pi \frac{u_I(t)}{L},
\ \ \
 a^I_2(x_1,x_2,t)=2\pi \frac{v_I(t)}{L}
 \label{auv}
\end{equation}
where $L$ is the size of the torus.
Substitute Eq. (\ref{auv}) into Eq. (\ref{LeffK}), we get
\begin{equation}
L=2\pi \t K_{IJ}  v_J\dot{u_I}
\end{equation}
which leads to the following commutator:
\be
\label{uvdel}
[u_I, v_J]= i(\t K^{-1})_{IJ} /2\pi
\ee
The large gauge transformation $\psi_a \to e^{i 2\pi n_{I} p^I_a x/L}\psi_a$
generate an equivalence relation
\be
\v u \sim \v u + \v n
\label{transu}
\ee
and $\v n$ satisfy
\be
n_{I} p^I_a ={\rm integer}
\ee
for all $a$. The vectors $\v n$ that satisfy the above condition form a
lattice whose basis vectors are denoted as $\v e_I$.
This lattice will be call the $\v e$-lattice.
The physically distinct $\v u$ points are all in the unit cell of the
$\v e$-lattice.

Since the conjugate variables $\v v$ also have the same
equivalence relation Eq. (\ref{transu}), the allowed value of $\v u$ are
quantized. To describe this quantization,
let us use the symmetric matrix $\t K$ to define an inner product
$\v u_1 \cdot \v u_2 \equiv u_{1I} \t K_{IJ} u_{2J}$ (which may not be
positive definite). Introduce a dual lattice (which will be called the $\v
d$-lattice) with the basis vectors ${\v d_I}$:
\be
\v d_I \cdot \v e_J = \delta_{IJ}
\ee
Then the allowed $\v u$'s all lie on the $\v d$-lattice, and
the ground states are labeled by the lattice points on the dual lattice
\cite{latt,WZ}.
However, the points connected by vectors in the $\v e$-lattice are gauge
equivalent:
\be
\v u \sim \v u + \v e_I, \ \ \ \v u \in \hbox{$\v d$-lattice}
\label{trans}
\ee
Thus only the $\v d$-lattice points which
lie inside the unit cell of the $\v e$-lattice
can represent independent ground states.

For ease of calculation, let us redefine the
gauge fields $a^I$ to make the basis of $\v e$-lattice to be the standard
basis vector ({\ie} $(\v e_I)_J=\del_{IJ}$). We will call such a basis the
primary basis. For the primary basis,
$p^I_a$ have the following two properties
\enu{
\item
$p^I_a$ are all integers.
\item
When viewed as $n$-dimensional vectors,
the vectors $\v p^I$ span a $\ka$ dimensional ``volume''
in the $n$-dimensional space.
This $\ka$ dimensional ``volume'' does not contain any $n$-dimensional
integer vectors. (Otherwise, we can choose a new set of $\v p^I|_{I=1..\ka}$
that span a smaller cube.)
}
{}From \Eq{Kqnu}, it is clear that $\t K$ is a symmetric integer matrix. This
$\t K$ matrix is similar to
the $K$ matrix in the $K$-matrix description of Abelian QH states
\cite{Kmat}.
For the primary basis,
the basis vectors of the $\v d$-lattice are given by the columns
of $\t K^{-1}$.
The number of the $\v d$-lattice points that lie inside the unit cell of the
$\v e$-lattice is given by $|$det$(\t K)|$.

For Abelian QH liquids, each point in the unit cell of the
$\v e$-lattice labels distinct ground state, and the ground state degeneracy is
given by $|$det$(\t K)|$ \cite{Wtop,abDeg}.
However, for non-Abelian states, there are additional
equivalent relations:\cite{WZ}
\be
\v u \sim T_i \v u, \ \ \ \v u \in \hbox{$\v d$-lattice}, \ \ i=1,2,...
\label{map}
\ee
where $T_i$ are linear maps which map a $\v d$-lattice point to another
$\v d$-lattice point.
Under the equivalence relation Eq. (\ref{map}) different $\v d$-lattice point
in the  unit cell of the
$\v e$-lattice can represent the same ground state. Thus only the
points in the folded unit cell represent distinct ground state.\cite{WZ}
Therefore, to calculate the ground state
degeneracy on torus, we need to  know $\t K$ and $T_i$.

Now let us describe how to get the maps $T_i$.
Recall that
in addition to the Abelian gauge transformations in $\cG_{abl}$,
there are additional discrete gauge transformations
$W_i$ as defined in Eq. (\ref{MpW}).
Since $W_i$ is a gauge
transformation, the physical states must satisfy $W_i|phys\>=|phys\>$.
The gauge transformations $W_i$ induces a gauge
transformation on $a^I_\mu$:
$a^I_\mu \to (T_i)_{IJ} a^J_\mu$, where $T_i$ is obtained from
Eq.  (\ref{MpW}).
Therefore $a^I$ and $(T_i)_{IJ} a^J$, hence $\v u$ and $T_i\v u$, are
equivalent points.

We would like to point out that the above result of ground state degeneracy on
torus is correct only when $\cG$ has no disconnected pieces. When $\cG$ has
disconnected parts, the above result needs to be modified. Let us assume
$\cG$ has a form $\cG=\cG_c\otimes\cG_d$ where $\cG_d$ is a discrete group.
Then the low energy effective theory is a ($\cG_c$ CS theory)$\times$($\cG_d$
gauge theory). The above calculation only calculates the ground states from
the  $\cG_c$ CS theory. We know that the discrete $\cG_d$ gauge theory
has $|\cG_d|^{2g}$ degenerate ground states on a genus $g$ surface, where
$|\cG_d|$ is the number of elements in $\cG_d$. Thus the total number of the
ground states is given by the number of ground states of the $\cG_c$ CS theory
times $|\cG_d|^{2g}$.

Before ending this section let us summarize the steps of
the projective construction as follow:
\enu{
\item
Introduce a few partons $\psi_a|_{a=1..n}$.
\item
Introduce a few electron operators
\be
\Psi^{(i)}_e= \sum_m C^{(i)}_m \prod_a \psi_{a}^{n_a^{(m)}}(z)
\label{PsiEpsi-gg}
\ee
(which generalizes \Eq{PsiEpsi-g}).
\item
Assign charge $Q_a$ to each parton such that the electron operators all have
charge $e$.
\item
Find the gauge group $\cG$ (see \Eq{gaugeT})
that leaves the electron operators $\Psi_e$ and
$Q$ unchanged (see \Eq{WQ}).
\item
Find the filling fractions $m_a$ which satisfy \Eq{WM} and \Eq{tMQ}.
}
This leads to a QH state with wave function
\begin{equation}
\Phi(\{ z^{(i)}_1,..., z^{(i)}_{N_i}\} )=
 \<0| \prod_i [\Psi_e^{(i)}(z^{(i)}_1)...
 \Psi_e^{(2)}(z^{(i)}_{N_i})] |\Phi_{parton}\> ,
\end{equation}
where $\Phi_{parton}$ is the free parton wave function in which the $a^{th}$
kind of partons form a $\nu=m_a$ QH state:
\begin{equation}
 \Phi_{parton} =\prod_a \chi_{m_a}(z^{(a)}_1,..., z^{(a)}_{N_a})
\end{equation}
Here $\chi_l$ is the fermion wave function with $l$ filled
Landau levels.
The bulk effective theory of the above state is given by
\Eq{eff-nab-g} (or \Eq{CS-g}). The edge effective theory is the
$U^{\sum|m_a|}(1)/\cG$ coset theory.
The filling fraction $\nu$ is given by $\nu=\sum_a m_a Q_a^2/e^2$.

To obtain the Abelian version of the effective theory,
we also need to do the following
\enu{
\item
Find a set of linearly independent integer vectors $\v p^I|_{I=1..\ka}$, such
that $P^I={\rm diag}(p^I_1,p^I_2,..., p^I_n)$ is in the Lie algebra of
$\cG_c$. We also require
$\v p^I$ to span a $\ka$ dimensional ``volume'' in the $n$ dimensional space
that does not contain any integer vectors.
}
Given $p^I$,
\Eq{WGW} and \Eq{MpW} determine the $T_i$ matrix.
\Eq{Kqnu} determines the $\t K$ matrix.
$T_i$, $\t K$ and $|\cG_d|$ allows us to determine the ground state
degeneracy on the torus.

\section{Applications of projective construction}

Now let us apply the above general results to some simple cases to gain
better understanding of the projective construction. First let us split an
electron into two partons
\be
\Psi_e= \psi_1\psi_2
\ee
with charges $Q_1=e\frac{l_1}{l_1+l_2}$ and $Q_2=e\frac{l_2}{l_1+l_2}$.
If $l_1\neq l_2$ the gauge group is $U(1)$:
$\cG=\{ e^{i\th \tau_3} \}$.
The effective theory is Eq. (\ref{eff-g}) with $\ka=1$ and $\v p^1 = (1,-1)$.
It is clear that $\v p^1$ forms a primary basis, since the line from
$(0,0)$ to $(1,-1)$ does not contain any integer points.
Let us consider a QH state in
which the two partons form integral
QH states with filling fraction $m_1$ and $m_2$. $m_a$ must
satisfy  \Eq{tMQ} (or Eq. (\ref{mQp})):
\be
m_1 l_1 - m_2 l_2 =0
\ee
Thus
\be
m_1=ml_2, \ \ \ m_2=ml_1
\ee
for an integer $m$.  The filling fraction of the QH liquid is
\ba
\nu &=& \frac{ml_2 l_1^2}{(l_1+l_2)^2} +  \frac{ml_1 l_2^2}{(l_1+l_2)^2}
\nonumber\\
&=& m \frac{l_1 l_2}{l_1+l_2}
\ea
The $\t K$-matrix is a 1 by 1 matrix:
\be
\t K = ml_2 + ml_1=m(l_1+l_2)
\ee
Since there is no additional discrete gauge transformations, the ground state
degeneracy on torus is $m(l_1+l_2)$.

When $m=l_2=1$ and $l_1=l$, we get a sequence of hierarchical state with
filling fraction $1/2$, $2/3$,..., $l/(l+1)$, ... (which are similar  to the
$1/2$, $2/5$,..., $l/(2l+1)$, ... states for the fermionic electrons).
The ground state degeneracies for those states are given by
$N_D=2,3,...,l+1,...$. Since the first parton has filling fraction $\nu=1$ and
the second parton has $\nu=l$, the electron wave function has a form
$\chi_1\chi_l$.

According to the $K$-matrix description,\cite{Kmat}
the effective theory of the $\chi_1\chi_l$ state is given by
\begin{equation}
\cL= \frac{ K_{IJ}}{4\pi} a_{I\mu}\pt_\nu a_{J\la} \eps_{\mu\nu\la}
\end{equation}
with $K=I_l+C_l$ where $I_l$ is the $l\times l$ identity matrix and $C_l$
is the  $l\times l$ matrix with all its elements equal to 1.
However, according to the projective construction, the $\chi_1\chi_l$ state
is describe by
\begin{equation}
\t \cL= \frac{ \t K_{IJ}}{4\pi} a_{I\mu}\pt_\nu a_{J\la} \eps_{\mu\nu\la}
\end{equation}
with $1\times 1$ matrix $\t K=(l+1)$.
Actually, there is no contradiction here. $\cL$ and $\t \cL$ are equivelant
topological theories (for example, they have the same number of degenerate
ground state).  $\t \cL$ can be regarded as a dual form of $\cL$.

We would like to remark here that showing the $\chi_1\chi_l$ state can be
described by a bulk effective CS theory with only one $U(1)$ gauge field
does not imply the  $\chi_1\chi_l$ state have only one branch of edge
excitations. Actually the  $\chi_1\chi_l$ state have $l$ branches of edge
excitations.

When $m=-l_2=1$ and $l_1=l$, we get a sequence of hierarchical state with
filling fraction $2$, $3/2$,..., $l/(l-1)$, ... (which are similar to the
$2/3$, $3/5$,..., $l/(2l-1)$, ... states for the fermionic electrons).
The ground state degeneracies for those states are given by
$N_D=1,2,...,l-1,...$

If $l_1=l_2=1$, the gauge group is $SU(2)$:
$\cG=\{ e^{i\v \th \cdot \v \tau} \}$. Now $m_1$ and $m_2$ must be equal
(see Eq. (\ref{WM})): $m_1=m_2=m$. The resulting state is nothing but the
$SU(2)_m$ non-Abelian state discussed in \Ref{Wnab} ($m=2$ case was discuss in
details in the section \ref{sec:su}). Its wave function is given by
$\chi_m^2$.
The filling fraction is $\nu=m/2$.

In the Abelian version of effective theory
Eq. (\ref{eff-g}),
we have $\ka=1$ and $\v q^1=(1,-1)$ as a primary basis. In addition to
the Abelian gauge transformation, we also have a discrete gauge transformation
$W=i\tau_2$. Such a discrete gauge transformation induces a ``gauge''
transformation on the Abelian gauge field $a^1_\mu \to -a^1_\mu $ (\ie $T=-1$
in Eq. (\ref{MpW}) ).  The $\t K$-matrix is
$\t K=2m$. The $2m$ $\v d$-lattice points in the unit cell of the
$\v e$-lattice are
$0, 1/2m$,..., $l/2m$,..., $(2m-1)/2m$. The $T=-1$ transformation leads to an
equivalence relation $l/2m \sim -l/2m \sim (2m-l)/2m$.
Thus the $SU(2)_m$ non-Abelian state has $m+1$ degenerate
ground states on torus.

Next we start with four different partons all with the same
charge $Q_a=1/2$.  The electron operator is chosen to be
\be
\Psi_e= \frac{1}{\sqrt{2}}
( \psi_{1}(z)\psi_{4}(z) -\psi_{3}(z)\psi_{2}(z) )
\label{PsiEpsi1}
\ee
[which is Eq. (\ref{Psiepsi}) if we identify
$(\psi_1,...,\psi_4)=(\psi_{1\up},\psi_{1\down},\psi_{2\up},\psi_{2\down})$].
The gauge group $\cG_c$ is
generated by 10 generators: $\tau_i \otimes \si_0$, $\tau_i\otimes\si_1$,
$\tau_i\otimes\si_2$, and
$\tau_0 \otimes \si_3$, where $\tau_0=\si_0$ are the 2 by 2 identity matrix.
It turns out that $\cG_c$ is the $SO(5)$ (or $Sp_4$)  group
in its 4 dimensional representation.
It appears that $\cG$ has no disconnected pieces and $\cG=\cG_c$.
To be consistent with the gauge invariance
Eq. (\ref{WM}), the partons must all have the same integer filling fraction
$\nu_a=m$.
The effective theory is
given by Eq. (\ref{eff-nab-g}) with $a_\mu$ in the Lie
algebra of the $SO(5)$ gauge group $\cG_c$.  After integrating out
the fermions , we get a $SO(5)_m$ CS theory.

The Abelian version of the effective theory Eq. (\ref{eff-g}) has $\ka=2$ and
\be
(p^1_1,...,p^1_4)=(1,0,0,-1), \ \ \ (p^2_1,...,p^2_4)=(0,-1,1,0)
\ee
The parallelogram spanned by $\v p^1$ and $\v p^2$ does not contain any integer
points. Thus $\v p^{1,2}$ is a primary basis.
The $\t K$-matrix is (see Eq. (\ref{Kqnu}))
$\t K=\pmatrix{2m&0\cr 0&2m\cr} $.
The $\v d$-lattice is generated by the basis $\v d^1 = (1/2m, 0)$ and
$\v d^2 = (0,1/2m)$. The $4m^2$ $\v d$-lattice points in the unit cell of
the $\v e$-lattice are $(k_1/2m,k_2/2m)$ with $k_1,k_2=0,...,2m-1$.
The electron operators is invariant under the following three transformations
\ba
W_1 &=& \pmatrix{
0&0&1&0\cr
0&0&0&1\cr
1&0&0&0\cr
0&1&0&0\cr
}  \nonumber\\
W_2 &=& \pmatrix{
0&0&0&1\cr
0&1&0&0\cr
0&0&1&0\cr
-1&0&0&0\cr
}  \nonumber\\
W_3 &=& \pmatrix{
1&0&0&0\cr
0&0&-1&0\cr
0&1&0&0\cr
0&0&0&1\cr
}
\ea
This induces three mappings on $\v u$ (see Eq. (\ref{map}))
\ba
T_1 &=& \pmatrix{
0&1\cr
1&0\cr
}  \nonumber\\
T_2 &=& \pmatrix{
-1&0\cr
0&1\cr
}  \nonumber\\
T_3 &=& \pmatrix{
1&0\cr
0&-1\cr
}
\ea
$T_{2,3}$ lead to equivalence relations $(k_1/2m,k_2/2m)\sim ((2m-k_1)/2m,
k_2/2m)\sim (k_1/2m, (2m-k_2)/2m)$.
Thus $k_1,k_2=0,...,m$ label all the independent states.
$T_1$ gives rise to an equivalence relation $(k_1/2m,k_2/2m) \sim
(k_2/2m,k_1/2m)$.
Therefore, the QH state obtained through the above projective construction
has $\frac{(m+1)^2}{2} + \frac{m+1}{2} = \frac{(m+1)(m+2)}{2}$
degenerate ground states
on a torus (represented by points $(k_1/2m,k_2/2m)$ with $k_1,k_2=0,...,m$
and $k_1\leq k_2$).

Note that when $m=1$ the above projective construction is just the
construction used in section \ref{sec:pf} to construct the filling
fraction $\nu=1$ bosonic Pfaffian state with wave function $\Phi^{pf}$
in Eq. (\ref{pf-Psie}).
We see that the $\nu=1$
bosonic Pfaffian state has 3 degenerate ground state on torus.
This result agrees with a previous result obtained from wave function.
\cite{pfDeg,RR} When $m>1$, the above construction produces new non-Abelian
states.

In  section \ref{sec:pf}, the bulk effective theory for the $\nu=1$ bosonic
Pfaffian state $\Phi^{pf}$ was found to be
the $U_{S_z}(1)\times SU_{color}(2)_2$ CS theory [see Eq. (\ref{pf-eff})].
{}From the above discussion, we see that he correct bulk
effective theory
should the $SO(5)_1$ CS theory.
However, it is not clear if the $U_{S_z}(1)\times SU_{color}(2)_2$ effective
CS theory is simply incorrect or it is equivalent to the $SO(5)_1$ CS theory.
On torus both theories give 3 degenerate ground states.
Also The edge excitation for the $\nu=1$
bosonic Pfaffian state should be described by the $U^4(1)/SO(5)_1$ coset theory.
Note that $U^4(1)=U(1)\times SU_{spin}(2)_2\times SU_{color}(2)_2$ theory
can be described by 8 free Majorana fermions. The $U(1)$ KM algebra
can be described by two Majorana fermions $(\Re \psi, \Im \psi)$,
the $SU_{spin}(2)_2$ KM algebra
by three Majorana fermions $(\eta_s^m |_{m=1,2,3})$, and the
$SU_{color}(2)_2$ KM algebra also by three Majorana
$(\eta_c^a |_{a=1,2,3})$. The $SO(5)$ gauge field couples to
$ (\eta_c^{1,2,3}, \eta_s^{1,2} )$ and the projection to the $SO(5)$ singlet
sector given us $U^4(1)/SO(5)_1=U(1)\times Ising$ theory described by
$(\psi, \eta_s^{3})$. Thus effective edge theory -- the $U(1)\times Ising$
theory -- obtained in  section \ref{sec:pf} is still valid.

We can also start with five different partons with four partons $\psi_{1,2,3,4}$
carrying charge $e l_1/2(l_2+l_2)$ and the fifth parton carrying charge
$e l_2/(l_1+l_2)$. The electron operator can be chosen to be
\be
\Psi_e= \frac{1}{\sqrt{2}}
( \psi_{1}(z)\psi_{4}(z) +\psi_{3}(z)\psi_{2}(z) )\psi_5
\label{PsiEpsi1-fpf}
\ee
(which is Eq. (\ref{Psipsi-fpf}) if we identify
$(\psi_1,...,\psi_5)=
(\psi_{1\up},\psi_{1\down},\psi_{2\up},\psi_{2\down},\psi_0)$).
The gauge group $\cG_c$ is
generated by 11 generators. The first 10 generators
$\tau_i \otimes \si_0$, $\tau_i\otimes\si_1$, $\tau_i\otimes\si_2$, and
$\tau_0 \otimes \si_3$ act only on $\psi_{1,2,3,4}$.
The last generator is given by ${\rm diag}(1,1,1,1,-2)$.
$\cG_c$ is the $SO(5)\times U(1)$ group.
Again $\cG$ has no disconnected pieces and $\cG=\cG_c$.
To be consistent with the gauge invariance
Eq. (\ref{WM}), the first four partons all have the same integer filling
fraction $\nu_a=m_1$ and the last parton $\psi_5$ has filling fraction $m_5$.
\Eq{tMQ} (or Eq. (\ref{mQp})) requires $m_1 l_1 - l_2 m_5=0$. Therefore
\be
m_1=ml_2,\ \ \
m_5=ml_1
\ee
The effective theory is
given by Eq. (\ref{eff-nab-g}) with $a_\mu$ in the Lie
algebra of the $SO(5)\times U(1)$ gauge group $\cG_c$.  After integrating out
the fermions , we get a $SO(5)_{m_1}\times U(1)$ CS theory.

The Abelian version of the effective theory is
Eq. (\ref{eff-g}) with
\ba
&&(p^1_1,...,p^1_5)=(1,0,0,-1,0), \nonumber\\
&&(p^2_1,...,p^2_5)=(0,-1,1,0,0), \nonumber\\
&&(p^3_1,...,p^3_5)=(0,1,0,1,-1)
\ea
The ``volume'' spanned by $\v p^{1,2,3}$ does not contain any integer
points and $\v p^{1,2,3}$ is a primary basis.
The $\t K$-matrix becomes
\be
\t K=\pmatrix{2m_1&0&-m_1\cr 0&2m_1&-m_1\cr -m_1&-m_1& 2m_1+m_5\cr }
\ee

The electron operators is invariant under the following three transformations
which also leave the Abelian gauge structure unchanged
\ba
W_1 &=& \pmatrix{
0&0&1&0&0\cr
0&0&0&1&0\cr
1&0&0&0&0\cr
0&1&0&0&0\cr
0&0&0&0&1\cr
}  \nonumber\\
W_2 &=& \pmatrix{
0&0&0&1&0\cr
0&1&0&0&0\cr
0&0&1&0&0\cr
-1&0&0&0&0\cr
0&0&0&0&1\cr
}  \nonumber\\
W_3 &=& \pmatrix{
1&0&0&0&0\cr
0&0&-1&0&0\cr
0&1&0&0&0\cr
0&0&0&1&0\cr
0&0&0&0&1\cr
}
\ea
This induces three mappings on $\v u$ (see Eq. (\ref{map}))
\ba
T_1 &=& \pmatrix{
0&1&0\cr
1&0&0\cr
0&0&1\cr
}  \nonumber\\
T_2 &=& \pmatrix{
-1&0&1\cr
0&1&0\cr
0&0&1\cr
}  \nonumber\\
T_3 &=& \pmatrix{
1&0&0\cr
0&-1&1\cr
0&0&1\cr
}
\ea

To calculate the ground state degeneracy on torus, let us consider only a
simple case $l_1=l_2=m=1$.
The corresponding HQ state is just the fermionic Pfaffian state at
 filling fraction
1/2 with wave function Eq.  (\ref{fpf-Psie}).
{}From $\t K^{-1}= \frac{1}{8} \pmatrix{5&1&2\cr 1&5&2\cr 2&2& 4\cr } $,
we find that the $\v d$-lattice is generated by the basis
$\v d^1 = (5/8, 1/8, 1/4)$,
$\v d^2 = (1/8, 5/8, 1/4)$,
and
$\v d^3 = (1/4, 1/4, 1/2)$.
The eight $\v d$-lattice points in the unit cell of the $\v e$-lattice are
$(5/8,1/8,1/4)$,
$(1/8, 5/8, 1/4)$,
$(7/8,3/8,3/4)$,
$(3/8,7/8,3/4)$,
$(0,0,0)$,
$(1/4, 1/4, 1/2)$
$(1/2,1/2,0)$,
and $(3/4,3/4,1/2)$.
$T_{2,3}$ do not lead to any new equivalence relations. However,
$T_1$ gives rise to two equivalence relations
$(5/8,1/8,1/4) \sim
(1/8,5/8,1/4) $ and
$(7/8,3/8,3/4) \sim
(3/8,7/8,3/4)$.
Thus the fermionic $\nu=1/2$ Pfaffian state has six degenerate states
on a torus.
This result again agrees with a previous result obtained from wave function.
\cite{pfDeg,RR}

\section{Summary}

In this paper we introduced a powerful method -- the projective construction
-- to construct many non-Abelian (and Abelian) states, which
including the fermionic $\nu=1/2$ and the bosonic $\nu=1$
Pfaffian states, and the $d$-wave paired non-Abelian state.
What is more significant is that
the projective construction allows us to calculate the bulk
and the edge effective theories. We find that the bluk effective theory
is a $SO(5)_1$ CS theory for the bosonic $\nu=1$ Pfaffian state, and
a $U(1)\times SO(5)_1$ CS theory for the fermionic $\nu=1/2$ Pfaffian state.
Using the bulk effective theory, the ground state degeneracy on torus is
calculated.

However, it is unclear if the projective construction can produce all the QH
states or not.  We still do not known how to use the projective
construction to construct the  Haldane-Rezayi state.\cite{HR}
Although we understand a lot of physical properties of the  Haldane-Rezayi
state,\cite{WWH,MiR,LW} we still do not know its bulk effective theory.

XGW is supported by NSF Grant No. DMR--97--14198 and by NSF-MRSEC Grant
No. DMR--94--00334. He wish to thank the warm hospitality and support from
National Center for Theoretical Sciences in Taiwan, where this work is
started.


\end{multicols}

\end{document}